\documentclass[letterpaper,superscriptaddress,nobibnotes,footinbib,aps,prb,twocolumn,showpacs,floatfix,longbibliography]{revtex4-1}

\usepackage[utf8]{inputenc}
\usepackage[T1]{fontenc}
 
\usepackage{amsmath,amssymb,amsfonts}
\usepackage{amstext, mathrsfs, textcomp}
\usepackage{nicefrac}
\usepackage{multirow}

\usepackage{graphicx}
\usepackage{xcolor}

\usepackage[colorlinks=true, citecolor=blue, urlcolor=blue, linkcolor=blue]{hyperref}

\usepackage{lipsum}

\usepackage{notes2bib}
\bibnotesetup{
note-name
 = ,
use-sort-key = false
}

\graphicspath{{figures/}{figures_appendix/}}

\newcommand{\TITLE}{Enhancement of electric and magnetic dipole transition of rare-earth doped thin films tailored by high-index dielectric nanostructures}

\hypersetup{
	pdftitle = {\TITLE},
	pdfsubject = {},
	pdfauthor = {Peter R. Wiecha et al.},
	pdfkeywords = {},
	pdfcreator = {LaTeX},
	pdfproducer = {LaTeX with hyperref and thumbpdf}
}

\begin{document}

\title{\TITLE}

\author{\firstname{Peter R.} \surname{Wiecha}}
\email[e-mail~: ]{peter.wiecha@cemes.fr}
\affiliation{CEMES, Universit\'e de Toulouse, CNRS, Toulouse, France}

\author{\firstname{Cl\'ement} \surname{Majorel}}
\affiliation{CEMES, Universit\'e de Toulouse, CNRS, Toulouse, France}

\author{\firstname{Christian} \surname{Girard}}
\affiliation{CEMES, Universit\'e de Toulouse, CNRS, Toulouse, France}

\author{\firstname{Arnaud} \surname{Arbouet}}
\affiliation{CEMES, Universit\'e de Toulouse, CNRS, Toulouse, France}

\author{\firstname{Bruno} \surname{Masenelli}}  
\affiliation{Universit\'e de Lyon, INSA-Lyon,  ECL,  UCBL,  CPE,  CNRS,  INL-UMR5270, Villeurbanne, France}

\author{\firstname{Olivier} \surname{Boisron}}  
\affiliation{Universit\'e de Lyon, CNRS, ILM-UMR 5306, Villeurbanne, France}

\author{Aurélie Lecestre}  
\affiliation{LAAS, Universit\'e de Toulouse, CNRS, INP, Toulouse, France}

\author{Guilhem Larrieu}  
\affiliation{LAAS, Universit\'e de Toulouse, CNRS, INP, Toulouse, France}

\author{\firstname{Vincent} \surname{Paillard}}
\affiliation{CEMES, Universit\'e de Toulouse, CNRS, Toulouse, France}

\author{\firstname{Aurélien} \surname{Cuche}}
\email[e-mail~: ]{cuche@cemes.fr}
\affiliation{CEMES, Universit\'e de Toulouse, CNRS, Toulouse, France}

\begin{abstract}
We propose a simple experimental technique to separately map the emission from electric and magnetic dipole transitions close to single dielectric nanostructures, using a few nanometer thin film of rare-earth ion doped clusters.
Rare-earth ions provide electric and magnetic dipole transitions of similar magnitude.
By recording the photoluminescence from the deposited layer excited by a focused laser beam, we are able to simultaneously map the electric and magnetic emission enhancement on individual nanostructures.
In spite of being a diffraction-limited far-field method with a spatial resolution of a few hundred nanometers, our approach appeals by its simplicity and high signal-to-noise ratio.
We demonstrate our technique at the example of single silicon nanorods and dimers, in which we find a significant separation of electric and magnetic near-field contributions.
Our method paves the way towards the efficient and rapid characterization of the electric and magnetic optical response of complex photonic nanostructures.
\end{abstract}


\maketitle


%
%
%


In the last decades, photonic nanostructures emerged as powerful instruments to control light at the subwavelength scale.\cite{novotny_principles_2006}
The interest in nano-optics lies usually in the control of the optical \textit{electric} field, since the response of materials to rapidly oscillating magnetic fields is extremely weak.
%
%
Actually, materials with a substantial magnetic response to electromagnetic radiation (\textit{i.e.} \(\mu\neq 1\)) are not known in nature.
However, properly designed nanostructures allow to significantly boost the magnetic response. 
For instance metallic (split-)ring resonators support a magnetic moment which is proportional to the area covered by the ring's aperture. 
For frequencies in the visible range this area is usually about \(10^6\) times larger than the equivalent area in atoms, defined by the Bohr radius, 
which explains the emergence of observable effects related to the optical magnetic field.\cite{giessen_glimpsing_2009}
In consequence, it is possible to overcome the natural limitation to \(\mu=1\) by designing so-called meta-materials, which are ordered arrangements of meta-units like split-ring resonators.\cite{pendry_negative_2000}

Using metals requires nanostructures of complex shape to obtain a significant magnetic response. 
On the other hand, in dielectrics of high refractive index, a magnetic response arises naturally from the curl of the electric field.\cite{merlin_metamaterials_2009, kuznetsov_magnetic_2012} 
Very simple geometries like spheres\cite{evlyukhin_demonstration_2012} or cylinders\cite{kapitanova_giant_2017} are actually sufficient to induce a strong magnetic field enhancement.\cite{kuznetsov_optically_2016} 
An additional advantage of high-index dielectric nanostructures is their low dissipation.
Silicon (Si) for example has a very low absorption through the entire visible range compared to noble metals.\cite{albella_electric_2014}
This weak absorption associated to the indirect gap in the near infrared becomes cumbersome only for applications involving propagation of visible light across distances of tens to hundreds of microns. 
Thus, in sub-micron dielectric nano-structures the absorption can usually be neglected.\cite{albella_low-loss_2013}
In summary, high-index nanostructures seem to provide an appropriate platform to study the confinement of optical electric and magnetic fields. 
Such nanostructures are indeed able to strongly localize those fields in different regions around them, while they are indistinguishable in the far field. In other words, the electric and magnetic parts of the photonic local density of states (LDOS) can be spatially separated around subwavelength small particles.\cite{rolly_promoting_2012, wiecha_decay_2018}

The separation of magnetic and electric field energy close to photonic nano-structures was first demonstrated by measuring the local magnetic field intensity using scanning near-field optical microscopy (SNOM). 
The sensitivity to the magnetic part of the optical near-field is provided by particular SNOM-tips, coated with nano-scale metal rings or split-ring resonators.\cite{devaux_detection_2000, devaux_local_2000, burresi_probing_2009, kabakova_imaging_2016} 
Another possibility to access the electric and magnetic near-field is to use optical transitions of quantum emitters and exploit the proportionality between their decay rate and the LDOS.\cite{girard_imaging_2004, aigouy_mapping_2014, carminati_electromagnetic_2015, cuche_near-field_2017}
However, to simultaneously obtain information on the electric and magnetic near-field components, probes supporting both electric (ED) and magnetic dipole (MD) transitions need to be used. 
This drastically limits the range of possible emitters: ED transitions are usually 4\,--\,5 orders of magnitude stronger than MD transitions.\cite{taminiau_quantifying_2012}
Lanthanoid ions (also known as rare earth elements) such as europium (Eu\(^{3+}\)) exhibit ED and MD transitions of similar strength\cite{carnall_spectral_1968} and can be used as probes of the optical magnetic near field.\cite{noginova_magnetic_2008, sanz-paz_enhancing_2018}
A peculiarity of Eu\(^{3+}\) is that electric and magnetic dipole transitions of comparable strength occur at close wavelengths. 
ED and MD transitions of Eu\(^{3+}\) both start from the \(^{5}D_{0}\) energy level via de-excitation to the \(^{7}F_{2}\) (ED) and the \(^{7}F_{1}\) level (MD).\cite{baranov_modifying_2017}
Those transitions lie in the visible, around \(\lambda_{\text{ED}}\approx 610\,\)nm and \(\lambda_{\text{MD}}\approx 590\,\)nm.\cite{noginova_magnetic_2008}
They lie at close wavelengths but are spectrally separated enough to be easily distinguished by using color filters.

\begin{figure}[t!]
	\begin{center}
		\includegraphics[page=1, scale=0.9]{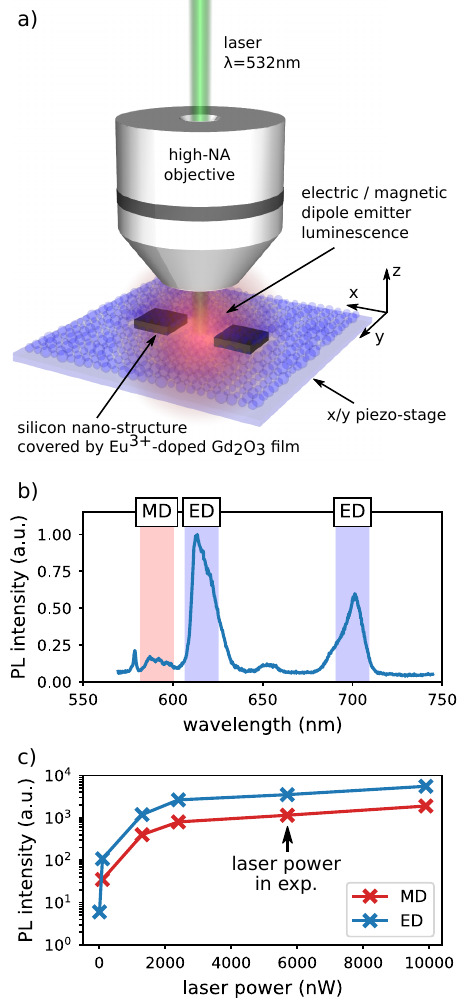}
		\caption{
			a) Sketch of the experimental setup. 
			A tightly focused laser beam (NA\,0.9, \(\lambda=532\)\,nm) is raster-scanned over a high-index dielectric nano-structure (in our case: silicon).
			An Eu\(^{3+}\)-doped Gd\(_2\)O\(_3\) film is deposited on the structure. 
			The photoluminescence due to the magnetic and electric dipole transitions of the  europium ions is collected in back-scattering by the microscope objective and analyzed with a spectrometer at each raster-scan position.
			b) Example spectrum of the photoluminescence from the Eu\(^{3+}\)-ion. Emission due to magnetic dipole (MD) and electric dipole (ED) transitions are underlined by red and blue color, respectively.
			c) PL intensity as function of laser power on the bare Eu\(^{3+}\) film (far from Si nanostructures). The power used for the acquisition of the maps was around \(5700\)\,nW.
		}\label{fig:setup}
	\end{center}
\end{figure}

%

%
Thanks to their unique properties, rare-earth ion doped media have been used as probes to determine the relative intensities of electric and magnetic LDOS at fixed positions or on non-structured samples like interfaces.\cite{karaveli_spectral_2011, hussain_spontaneous_2014, choi_selective_2016}
By attaching an Eu\(^{3+}\) doped nano-crystal to a SNOM tip, the \(3\)D spatial distribution of the electric and magnetic LDOS on top of gold strips has been recorded.\cite{aigouy_mapping_2014}
The methods employed in these studies are either based on rare-earth emitters at fixed position,\cite{noginova_magnetic_2008, choi_selective_2016} or require complex experimental setups (SNOM-type approach).\cite{aigouy_mapping_2014, cuche_fluorescent_2009}

In this article, we propose an alternative and complementary approach: 
we deposit a homogeneous \(10\,\)nm thin film of Eu\(^{3+}\)-doped nano-clusters on high-index dielectric nanostructures and raster-scan the sample under a tightly focused diffraction limited laser beam.
In this configuration, the photoluminescence maps probe the spatial variations of the ED and MD emissions around the silicon nanostructures. 
We compare the results with computed maps of the electric and magnetic radiative LDOS. 
The results show that our far field set-up allows a fast scanning with a good signal-to-noise ratio and a good spatial resolution. 
Interestingly, the far field PL shows a striking similarity with the radiative LDOS.

\section{Photoluminescence from Eu\(^{3+}\)-doped nanocluster film deposited on silicon nanostructures}

\subsection*{Sample preparation}

Our sample contains single crystalline silicon dimers consisting of two elements, each \(300\times 300 \times 90\,\)nm\(^3\) (\(L\times W\times H\)) large, separated by a gap \(G\) of variable size.
These nanostructures are fabricated in a top-down approach, where a single layer of negative-tone resist, namely hydrogen silsesquioxane (HSQ), is patterned by electron-beam lithography.\cite{guerfi_high_2013}
Subsequent reactive ion etching in the \(H = 90\)\,nm Si overlayer of silicon on insulator (SOI) substrates defines the structures. 
The remaining HSQ-layer on the top of the structures induces an additional SiO\(_2\) capping of approximately 20nm.
This layer acts as a spacer between the nanostructures and the Eu\(^{3+}\) doped film.
For more details on the fabrication process, see Ref.~\onlinecite{wiecha_evolutionary_2017}.
After the lithography, a \(30\,\)nm thick film of Eu\(^{3+}\)-doped Gd\(_2\)O\(_3\) nano-clusters is deposited.
This thickness is a good trade-off between luminescence intensity from the emitters and planar homogeneity (see also Appendix~\ref{appendix:backgroud}).
The film is synthesized by the Low Energy Cluster Beam Deposition (LECBD) technique.
LECBD consists in the ablation of a solid Gd\(_2\)O\(_3\) pellet, doped at 7\% with europium, \textit{via} a pulsed Nd:YAG laser (\(10\)\,ns pulse width). 
The chosen doping concentration results in a good compromise to preserve the stoichiometry of the sesquioxyde matrix as well as a high luminescence of the Europium ions, without impairing the crystallographic and optical properties of the clusters.\cite{masenelli_shells_2012}
The ablated fragments are then broken into small clusters, first by the injection of He (\(20\)\,mbar) as a buffer gas in the nucleation chamber and subsequently during its adiabatic expansion through a micrometer nozzle.
The resulting nanocrystals are deposited on the dielectric nanostructures without further breaking (``soft landing'').
More details about the process and the resulting nanocrystals can be found in Refs.~\onlinecite{cuche_fluorescent_2009, perez_functional_2010}.

\subsection*{Experimental configuration}

\begin{figure}[t!]
	\begin{center}
		\includegraphics[page=1]{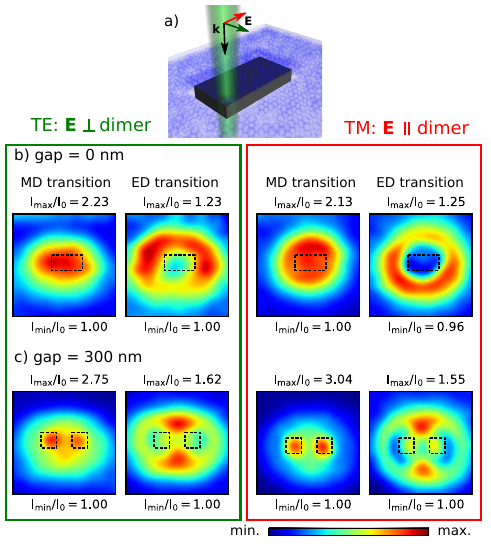}
		\caption{
			Experimental mappings for different laser polarizations (sketch shown in a) for gap sizes of b) \(G=0\,\)nm and c) \(G=300\,\)nm.
			left: TE polarized laser (electric field perpendicular to dimer long axis)
			right: TM polarized laser (\(\mathbf{E}\) along dimer long axis)
		}\label{fig:rasterscan_incident_polar}
	\end{center}
\end{figure}

\begin{figure*}[t!]
	\begin{center}
		\includegraphics[page=1]{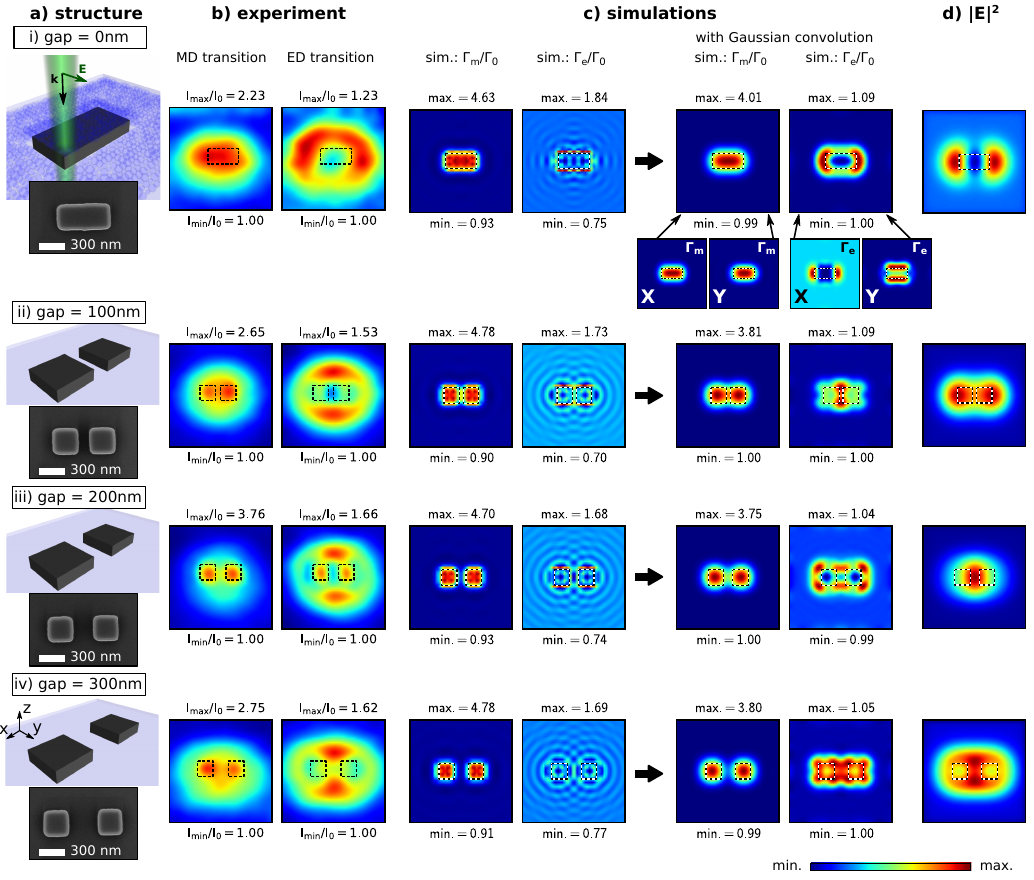}
		\caption{
			Comparison of the experimental results with simulated decay-rate maps for different gap sizes
			i) \(G=0\)\,nm, ii) \(G=100\,\)nm, iii) \(G=200\,\)nm and iv) \(G=300\,\)nm.
			a) Sketches and SEM images of the corresponding structures. SEM images were taken prior to Eu\(^{3+}\)-cluster deposition. Scale bars are \(300\,\)nm.
			b) Experimental results recorded with an incident polarization along the dimer axis. 
			Left column: MD transition, right column: ED transition.
			c) Simulated decay-rate maps before (left columns) and after convolution with a Gaussian profile (right columns).
			The subplots on the left show the magnetic dipole in-plane decay rates \(\Gamma_{m}^{xy}\) (\(XY\) averaged), the subplots on the right the respective ED maps \(\Gamma_{e}^{xy}\).
			The waist of the Gaussian profile is \(\eta=200\,\)nm.
			Insets in i)-c) show the (Gaussian convoluted) LDOS maps for \(X\)- and \(Y\)-oriented dipoles only.
			All color-maps show areas of \(2\times 2\)\,\textmu m\(^2\), dashed lines indicate the positions of the silicon dimer.
			d) simulated maps of the average electric field intensity in the Eu\(^{3+}\) film as function of the spot position for a diffraction limited, TE polarized laser (\(\lambda=532\,\)nm, FWHM of \(300\,\)nm).
		}\label{fig:dimer_exp_sim}
	\end{center}
\end{figure*}

Our experimental setup consists of an optical microscope equipped for spectroscopy with a confocal pinhole of 100\,\textmu m.
A simplified sketch of the experiment is shown in figure~\ref{fig:setup}a.
A linearly polarized cw laser at \(\lambda=532\,\)nm is focused by a \(\times 100\) air objective (\(\text{NA}=0.9\)) on the sample, which lies on a \(XY\) piezo stage. 
The photoluminescence (PL) from the sample is collected by the same microscope objective. 
The PL is dispersed by a grating (\(300\) grooves per mm) and detected by a CCD camera (Andor iDus 401), with an integration time of \(t_{\text{exposure}}=2\,\)s. 
A typical spectrum and saturation curve from the Eu\(^{3+}\)-doped Gd\(_2\)O\(_3\) film is shown in figure~\ref{fig:setup}b-c.
Finally, we obtain 2D PL maps by raster-scanning the sample under the laser with the piezo-stage.
At each position, we acquire a spectrum and extract the emission intensities from the spectral ranges corresponding to magnetic or electric transitions (\(\lambda=(590 \pm 10)\,\)nm, respectively \(\lambda=(610 \pm 10)\,\)nm) during post-processing.
The signal is normalized to the background PL intensity, recorded far from the nanostructures.
We note that our approach yields a high signal-to-noise ratio, which can be qualitatively seen for example in the spectra (see Fig.~\ref{fig:setup}b, and experimental maps in Figs.~\ref{fig:rasterscan_incident_polar} and~\ref{fig:dimer_exp_sim}).

\subsection*{Experimental results}

Figure~\ref{fig:rasterscan_incident_polar} shows raster-scan maps, measured on silicon nanorod dimers with gap-sizes of \(0\,\)nm and \(300\,\)nm, obtained with an incident laser polarization perpendicular to (TE) and along (TM) the dimer long axis.
In the experimental maps, a clear inversion of the contrast between emission from MD and ED transitions is observed for both polarizations. 
The MD PL yields a maximum intensity when the laser beam is focused on the Si blocks of the structure. 
On the other hand, the ED emission maps contain valley-like features with a reduced PL on top the Si blocks and an increased PL around these blocks.

Interestingly, both TE and TM configurations yield almost identical PL maps. This could be surprising at first sight as the spatial near field distribution around the Si nanodimers depends significantly on the laser polarization (see Appendix~\ref{appendix:excitation_field}) and at our laser wavelength, ED and MD transitions in Eu\(^{3+}\) are both populated by excitation of the \(^5\)D\(_1\) level via the local \textit{electric} field.\cite{baranov_modifying_2017, kasperczyk_excitation_2015}
Furthermore, it appears that the emitters are excited close to saturation, since in Fig.~\ref{fig:setup}c a plateau in the PL intensity occurs.
Both considerations about the laser polarization intensity and the excitation at saturation lead to the conclusion that the observed variations in the ED (resp. MD) PL maps are predominantly driven by  the electric (resp. magnetic) radiative LDOS at the \textit{emission} wavelength. 
This conclusion is in agreement with a recent study of the directional emission of Eu\(^{3+}\) ions, coupled to plasmonic nanostructures.
In the latter, the most significant modification of the Eu\(^{3+}\)-PL is ascribed to the improved light out-coupling, whereas the plasmonic enhancement of the excitation has only a weak impact on the emission.\cite{murai_directional_2017} 

We systematically measured PL maps on silicon dimers with gap-sizes between \(0\,\)nm and \(300\,\)nm, which are shown in figure~\ref{fig:dimer_exp_sim}b.
For these results, the laser polarization was perpendicular to the long axis of the dimers.
Sketches and scanning electron microscopy (SEM) images of the corresponding structures are shown in Fig.~\ref{fig:dimer_exp_sim}a.
As already observed in figure~\ref{fig:rasterscan_incident_polar}, a clear inversion of the contrast between emission from MD and ED transitions is observed. 
A single nano-rod (figure~\ref{fig:dimer_exp_sim}i) yields one peak (MD) or one valley feature (ED). 
However, as soon as a gap is introduced (figures~\ref{fig:dimer_exp_sim}ii-iv), the maps start to change significantly. 
The magnetic dipole emission yields confined hot-spots at the silicon block positions,
whereas the PL maps from the ED transition show hot-spots outside the dimer, located above and below the gap. 


%

\begin{figure}[t!]
	\begin{center}
		\includegraphics[page=1]{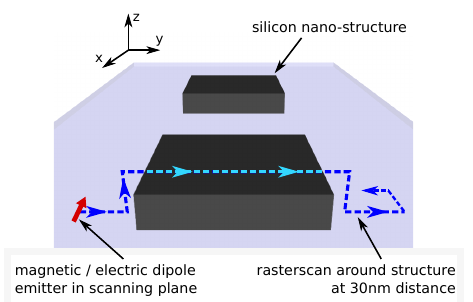}
		\caption{
			Sketch of the simulation procedure. 
			A magnetic or electric dipolar emitter is raster-scanned across the nano-structure at \(30\,\)nm above its surface, simulating the distance of the doped film to the top surface of the structures (\(20\,\)nm SiO\(_2\) spacing layer). 
			Far from the structure, the scan-height is at \(30\)\,nm above the ground-level.
			The raster-scan is repeated for emitter orientations along \(0X\) and \(0Y\), the decay-rates are averaged.
		}\label{fig:simulation}
	\end{center}
\end{figure}

\section{Calculating the electric and magnetic radiative LDOS}

To elucidate the origin of the strong difference in recorded ED/MD emission maps, we perform in the following simulations of the electric and magnetic LDOS as well as of the near-field distribution at the excitation wavelength, which are the two main factors, driving the photoluminescence of the emitters.

We will make the assumption that the angular emission pattern cannot explain the inversion of contrast, since we always collect the average emission from a large number of emitters with random orientations.


\subsection*{Decay-rate of electric/magnetic emitters}

The decay-rate enhancement of a quantum emitter is proportional to the partial LDOS at the emitter's location.\cite{carminati_electromagnetic_2015,lunnemann_local_2016}
With ``partial'' LDOS, we refer to the projection of the LDOS onto a specific direction (dipole orientation) and to the nature of the dipole transition (electric or magnetic).
The full LDOS can be obtained by averaging over all possible dipole orientations. 
Hence, the computation of the decay-rate modification of electric and magnetic dipole transitions also yields the photonic local density of states.

The decay-rate of electric or magnetic dipolar transitions is modified by the presence of a polarizable material. 
The effect is intuitively understandable for an electric dipole transition \(\mathbf{p}\), as a result of the enhancement (or weakening) of the electric near-field due to the dielectric contrast or optical resonances, and the back-action of the radiated field on the dipole emitter.
It is possible to derive an integral equation describing the decay-rate \(\Gamma_{e}\) of an electric dipole transition at a position \(\mathbf{r}_0\), close to an arbitrary nano-structure:\cite{carminati_electromagnetic_2015,wiecha_decay_2018}
\begin{multline}\label{eq:gamma_electric}
\Gamma_{e}({\bf r}_{0},\omega) = \Gamma_{e}^{0}(\omega)
\\ 
\times \left(1+
\frac{3}{2k_{0}^{3}}{\bf u}\cdot \text{Im}\big(
{\cal S}_{p}^\text{EE}({\bf r}_{0},{\bf r}_{0},\omega)\big)\cdot{\bf u}
\right) \; ,
\end{multline} 
In this equation, \(\Gamma_{e}^{0}(\omega)$ = $ 4k_{0}^{3} p^{2}/3\hbar\) is the decay rate of the emitter in the absence of a structure with the vacuum wavenumber \(k_0\). ${\bf u}$ and \(p\) denote the dipole orientation and amplitude, respectively,  and
\begin{multline}\label{eq:gamma_magnetic_SpEE}
{\cal S}_{p}^\text{EE}({\bf r},{\bf r}_{0},\omega) =
\int\limits_{V} \text{d}{\bf r}'\int\limits_{V} \text{d}{\bf r}''\chi({\bf r}',\omega)\mathbf{G}^\text{EE}({\bf r},{\bf r}',\omega)
\\
\cdot\, \mathbf{K}({\bf r}',{\bf r}'',\omega)\cdot\mathbf{G}^\text{EE}({\bf r}'',{\bf r}_{0},\omega)
\; .
\end{multline}
\(\mathbf{K}\) is the generalized propagator (see \textit{e.g.} Ref.~\onlinecite{martin_generalized_1995}).
The propagator \(\mathbf{G}^\text{EE}\) can be found by identification using the electric field emitted by a dipolar source\cite{agarwal_quantum_1975} and the Green's Dyad for vacuum\cite{martin_generalized_1995} (\textit{c.f.} also Ref.~\onlinecite{wiecha_decay_2018}).

Eq.~\eqref{eq:gamma_electric} describes the decay-rate of an \textit{electric} quantum emitter. 
It turns out, that although no known material has a direct response to rapidly oscillating magnetic fields, the decay rate of a magnetic dipole transition is nevertheless influenced by the presence of material.
Such \textit{magnetic-magnetic} response function associated with a structure which, by itself, has no direct magnetic response, arises from the \textit{electric} field emitted by the magnetic dipole, which can interact with the environment. 
In particular, Mie-type optical resonances in dielectric nanostructures often induce curled features in the electric near-field spatial distribution, leading to a strong enhancement of the magnetic field via the Maxwell's equation \(\text{i}k_0 \mathbf{B}(\mathbf{r}) = \text{rot}\,\mathbf{E}(\mathbf{r})\) (for monochromatic fields, \(k_0=2\pi/\lambda_0\)). 
Formally, the magnetic decay-rate in the vicinity of arbitrary nanostructures can be calculated in complete analogy to equation~\eqref{eq:gamma_electric} by replacing the tensors \(\mathbf{G}^\text{EE}\) in equation~\eqref{eq:gamma_magnetic_SpEE} with the Dyads \(\mathbf{G}^\text{HE}\) (first occurrence) and \(\mathbf{G}^\text{EH}\) (second occurrence, \textit{c.f.} reference~\onlinecite{wiecha_decay_2018}). 
The latter tensors are often called ``mixed-field susceptibilities''.\cite{girard_optical_1997, schroter_modelling_2003, sersic_magnetoelectric_2011}
%


\subsection*{Simulation of raster-scan maps}

To solve Eq.~\eqref{eq:gamma_electric} numerically, the integrals in Eq.~\eqref{eq:gamma_magnetic_SpEE} can be converted to discrete sums over the mesh-points of a volume-discretized nanostructure.
To calculate raster-scans of the MD and ED decay-rates (hence maps of the electric and magnetic LDOS), we make use of the concept of a generalized propagator.\cite{martin_generalized_1995}
This approach significantly speeds up the computation of the decay-rate at multiple locations, as explained in reference~\onlinecite{teulle_scanning_2012}.

The specific simulation procedure is illustrated in figure~\ref{fig:simulation}.
We raster-scan a dipolar emitter of either magnetic (\(\Gamma_m\), \(\lambda_{\text{MD}}=590\,\)nm) or electric nature (\(\Gamma_e\), \(\lambda_{\text{ED}}=610\,\)nm) across a discretized nano-dimer (discretization step \(s=20\,\)nm).
The dimer lies in a homogeneous environment of refractive index \(n_{\text{env}}=1.8\), corresponding to the optical index of the Gd\(_2\)O\(_3\) film.\cite{dakhel_optical_2001} 
The distance between emitter and structure is kept at \(30\,\)nm. 
Far from the structure, the dipole is scanned at \(30\,\)nm above the substrate. 
In this way, we account for the distance between the Eu\(^{3+}\)-cluster film and the nanostructures due to the 20 nm thick SiO\(_2\) layer.
%
With a finite numerical aperture as used in our experiment, we expect that only a very small part of the radiation from dipole emitters oriented along \(OZ\) will reach our detector (due to their toroidal radiation pattern propagating perpendicular to the \(Z\)-axis). 
We show indeed in Appendix~\ref{appendix:zdipoles}, that a small fraction of contributing \(Z\)-dipoles does not strongly impact on the LDOS maps.
Therefore, we will neglect dipole emitters oriented along \(OZ\) in the simulations shown in figure~\ref{fig:dimer_exp_sim}.
%
%
Thus, the simulated maps only take into account dipoles located in a plane parallel to the substrate (the computed values being the average of the contributions of dipoles oriented along \(OX\) and \(OY\)).
We denote these average decay rates \(\Gamma_m^{xy}\) and \(\Gamma_e^{xy}\) for MD and ED, respectively. 

The direct result of this procedure has a far too high spatial resolution (see the two columns on the left of figure~\ref{fig:dimer_exp_sim}c).
Therefore, to account for the diffraction limited size of the laser-spot, and hence the excitation of a whole ensemble of emitters, we convolve the simulated mapping with a Gaussian profile of waist \(\eta=200\)\,nm (corresponding to a full width at half maximum of around 450\,nm which is an upper estimate for the emission area.\cite{wertz_single-molecule_2015,mack_decoupling_2017-1} 

Our numerical signal power is described by the following equation:
\begin{equation}\label{eq:raster_scan_intensity}
\begin{aligned}
I_{\text{signal}}(\mathbf{r}) \
= & 
\int\limits_{A_{\text{spot}}}  \hbar\omega_0\, n(\mathbf{r}')\, \Gamma_i^{xy}(\mathbf{r}') \exp\left( -\frac{\mathbf{r}'^2}{2\eta^2} \right)\, \text{d}{\mathbf{r}}'
\, .
\end{aligned}
\end{equation}
The integral runs over the area of the laser-spot \(A_{\text{spot}}\), \(\hbar \omega_0\) is the photon energy of the MD or ED transition, \(n\) is the density of emitting Eu\(^{3+}\) ions (which we consider homogeneous). 
\(\Gamma_i^{xy}\) is the \(x\)-\(y\)-averaged decay rate of either a magnetic (\(i=\ \)``\(m\)'') or an electric (\(i=\ \)``\(e\)'') dipole transition. 
The exponential factor accounts for the Gaussian intensity profile of the laser beam.
The simulated maps after convolution are shown in the right columns of Fig.~\ref{fig:dimer_exp_sim}c.
We note that a similar procedure has been successfully used to recover the near-field intensity distribution from two-photon luminescence (TPL) measurements on plasmonic nanostructures.\cite{ghenuche_spectroscopic_2008}

\subsection*{Discussion}

Comparing the simulated and experimental results in figure~\ref{fig:dimer_exp_sim}, we observe that the features in the simulated maps are more confined around the dimer blocks. 
Nevertheless there is a general qualitative agreement. The global features and trends observed in the experimental maps are reproduced by the simulations.
The contrasts of electric and/or magnetic LDOS close to dielectric nanostructures are in agreement with recently published experimental results, showing a clear separation of electric and magnetic LDOS above the dielectric structures.\cite{makarov_nanoscale_2018, sanz-paz_enhancing_2018}
Also the quantitative trends in the intensity ratios are correctly recovered by the simulations, showing an enhancement of the luminescence intensity by a factor of around 3 for the MD transition and 1.5 in the ED case.

Compared to the simulations, the experimental exhibit broader features.
Several effects can contribute to this broadening, first of all, not all emitters are excited at saturation power.
The laser spot is of finite, diffraction limited size, hence excites a large collection of emitters at each raster-scan position.
Those at the edge of the laser focal spot are excited with a weaker field amplitude, hence not at saturation.
Therefore, a residual contribution of the excitation electric field at the laser wavelength cannot be totally neglected.
In the LDOS simulations on the other hand, we only take into account a single, saturated point-emitter, raster-scanned across the silicon dimer.
To elucidate this effect, we calculate the average electric near-field intensity in the Eu\(^{3+}\) doped film at each laser-spot position, shown in figure~\ref{fig:dimer_exp_sim}d for TE polarization (see also Appendix~\ref{appendix:excitation_field} for the TM case).
In these calculations, we perform a numerical raster-scan with a focused illumination at \(\lambda=532\,\)nm and with a spot size of 300\,nm FWHM, corresponding to our experimental configuration. At each position of the focused illumination, we calculate the average total field intensity in the entire Eu\(^{3+}\) doped film.
There seems to be a certain resemblance between electric LDOS and excitation field pattern, which makes sense, because the near-field is correlated to the LDOS, even though we compare LDOS and \(|\mathbf{E}|^2\) at different wavelengths. 
The magnetic part of the LDOS on the other hand does not show comparable features.
This clear contrast suggests, that close to saturation, the experimental mappings are mainly driven by the electric and magnetic LDOS, while a certain broadening  and modification can be explained by a residual contribution of the excitation electric field of the laser.
In addition, light might be guided through the dimer when the laser hits one extremity, and Europium ions at the other end, far from the laser spot can be remotely excited in this case as well. 
Such effect can result in a significant broadening of the spatial features in our mappings,\cite{wertz_single-molecule_2015,mack_decoupling_2017-1} and is not fully taken into account in our current theoretical model, even if an enlarged emission area is considered via the Gaussian convolution.

A better agreement could probably be achieved by simulating an actual ``layer'' of dipoles, instead of raster-scanning a single emitter, and by taking into account the modified global radiation pattern, since the angular emission of single magnetic and electric dipoles can be strongly affected by the inhomogeneous environmentt.\cite{curto_multipolar_2013, colas_des_francs_relationship_2001, alaee_generalized_2015}
However, this would require to consider the excitation intensity as function of the emitter position with respect to the center of the focal spot. 
A possible approach to the problem of non-saturated emitters could be the description of the Eu\(^{3+}\) emitters in the layer as three-level molecules.\cite{girard_generalized_2005, colas_des_francs_theory_2007}
These more sophisticated theoretical approaches lie outside the scope of this study and will be the subject of a future work.

%

\section{Conclusion}

In conclusion, we demonstrated that the enhancement of magnetic and electric dipole transitions is spatially tailored in the near-field of dielectric nanostructures. 
We obtained our results with a technique which allows the parallel mapping of the electric and magnetic components of the photoluminescence of an Eu\(^{3+}\) doped film in a far-field detection scheme.
To this end, a nanometer thin film of rare-earth ion doped clusters is deposited on top of the structures and excited by a tightly focused laser beam. The laser beam is raster-scanned across the structures to obtain \(2\)D maps of the ED and MD emission enhancement.
Numerical simulations of the electric and magnetic LDOS show a general qualitative agreement with the experimental results and provide a good understanding of the underlying physical effects.
These results show that the electric and magnetic near-field intensity is confined in distinct regions around simple dielectric nanostructures such as rods or dimers.
Our work paves the way towards the very rapid and simple experimental characterization of the response of photonic nanostructures to both, the electric and magnetic field of visible light and towards the design of ``on demand'' magnetic and electric near-field landscapes by tailoring the resonances (like Mie-type modes, bound states in the continuum) of dielectric nanostructures.\cite{he_toroidal_2018}

\section{Appendix -- Supplemental Material}
\subsection{Excitation field distribution for different laser polarizations}\label{appendix:excitation_field}
\begin{figure}[h!]
    \begin{center}
        \includegraphics[page=1, width=0.8\columnwidth]{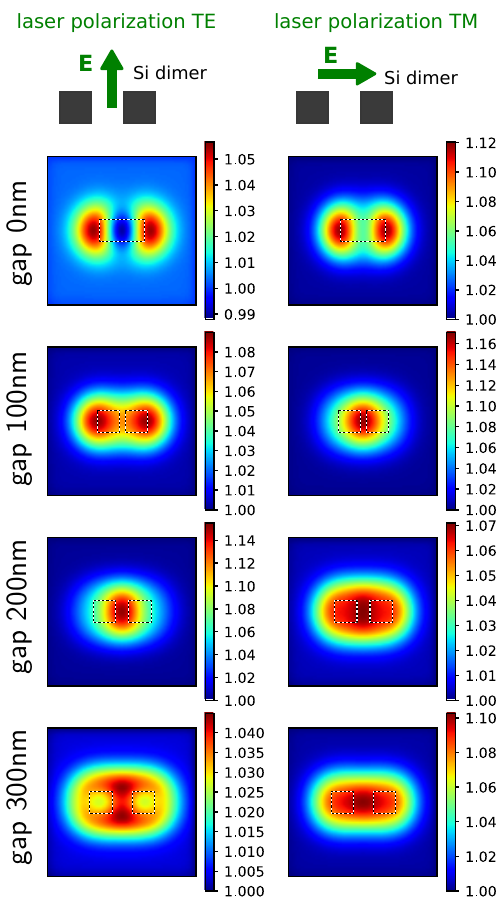}
        \caption{
            Fundamental field distribution as function of laser position for TE (left column) and TM (right column) laser polarization. Calculated at \(\lambda_{\text{laser}} = 532\,\)nm. 
        }\label{fig:fundmental_E_TE_TM}
    \end{center}
\end{figure}

To assess how the excitation field in the europium film depends on the focal position of the laser, we simulated in figure~\ref{fig:fundmental_E_TE_TM} the average electric field intensity \(|\mathbf{\overline{E}}|^2\) induced by the laser-beam, averaged over the whole depot-layer.
Each pixel corresponds to the average near-field intensity in the entire film, if the laser is focused at the according position. 
The left column shows the case of laser light polarized perpendicular to the dimer axis (TE), while the right column depicts the case of a laser polarization along the dimer axis (TM).
From top to bottom, increasing gap sizes of \(0\,\)nm, \(100\,\)nm, \(200\,\)nm and \(300\,\)nm are shown.
Obviously, the field distribution as function of the laser focus position is dependent on the polarization.
Together with the observation that the PL-mappings are independent of the laser polarization (see main text figure~\ref{fig:rasterscan_incident_polar}), we conclude that most of the Eu\(^{3+}\) ions in the laser focal spot are excited at saturation.

\subsection{Influence of \(Z\)-oriented dipole transitions}\label{appendix:zdipoles}
\begin{figure}[h!]
    \begin{center}
        \includegraphics[width=0.9\columnwidth]{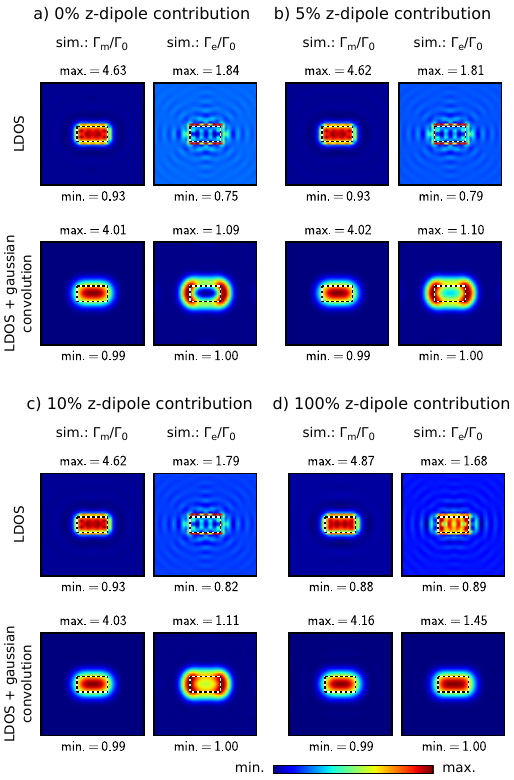}
        \caption{
            LDOS maps above a silicon nanorod with increasing contributions of \(Z\)-oriented dipoles. (a) only \(X\) and \(Y\) dipoles, (b) 5\%, (c) 10\% and (d) 100\% contributions of \(Z\) dipoles.
        }\label{fig:influence_z_dipoles}
    \end{center}
\end{figure}

To assess the influence of dipoles along \(z\) (perpendicular to the scanning plane), we simulated LDOS maps with gradually increased contribution of these emitters. 
Fig.~\ref{fig:influence_z_dipoles}a) shows the LDOS for only \(x\) and \(y\) dipole orientations. In b) \(z\)-dipoles contribute by \(5\,\%\) of their magnitude, in c) by \(10\,\%\) and in d) \(z\)-dipoles contribute fully.
The decay rate of dipoles along \(z\) experiences the strongest influence through the presence of the dielectric structure, with a very narrow confinement of the LDOS in both cases (for ED and MD).
Thus the mappings change significantly, as soon as signal from the perpendicular dipoles is recorded. 
Hence, we conclude that the dipoles with \(z\)-orientation are almost invisible in our detection scheme, since their radiation lobe lies in the \(xy\) plane.

\subsection{Background PL}\label{appendix:backgroud}
\begin{figure}[h!]
    \begin{center}
        \includegraphics[width=0.9\columnwidth]{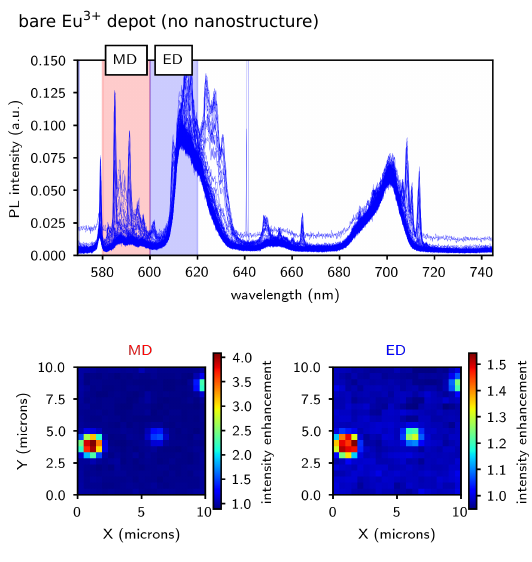}
        \caption{PL spectra on the bare Eu\(^{3+}\)-doped film (deposited on the blank substrate, \textit{i.e.} without any nanostructure). Top: superposition of all spectra recorded in the raster-scan and bottom: intensity maps (left subplot: MD, right subplot: ED). 
        }\label{fig:background_PL}
    \end{center}
\end{figure}

Photoluminescence measurements on the bare film (i.e. in absence of any silicon nanostructures) are shown in figure~\ref{fig:background_PL}.
The PL is very homogeneous over large areas of many microns, with only some defects. 
These defects do not disturb the measurements, because they are significantly larger than the spatial features obtained from scanning the silicon nanostructures, and therefore can be easily identified.

\begin{acknowledgments}
    This work was supported by Programme Investissements d'Avenir under the program ANR-11-IDEX-0002-02, reference ANR-10-LABX-0037-NEXT, by the LAAS-CNRS micro and nanotechnologies platform, a member of the French RENATECH network and by the computing facility center CALMIP of the University of Toulouse under grant P12167.
\end{acknowledgments}

\end{document}